\documentclass[aoas,nameyear,dvips]{arximspdf}
\usepackage{dcolumn}
\usepackage{graphicx}

\doi{10.1214/09-AOAS301}
\volume{4}
\issue{2}
\pubyear{2010}
\firstpage{791}
\lastpage{804}

\makeatletter
\newcolumntype{d}[1]{D{.}{.}{#1}}
\makeatother

\begin{document}
\begin{frontmatter}

\title{Building a model for scoring 20 or more runs in a~baseball game}
\runtitle{Modeling rare baseball events}

\begin{aug}
\author[A]{\fnms{Michael R.} \snm{Huber}\ead[label=e1]{huber@muhlenberg.edu}\corref{}}
and
\author[B]{\fnms{Rodney X.} \snm{Sturdivant}}
\runauthor{M. R. Huber and R. X. Sturdivant}
\affiliation{Muhlenberg College and United States Military Academy}
\address[A]{Department of Mathematics\\
\quad and Computer Science\\
Muhlenberg College\\
Allentown, Pennsylvania 18104\\
USA\\
\printead{e1}} %adresu isvedimo komanda gale!
\address[B]{Department of Mathematical Sciences\\
United States Military Academy\\
West Point, New York 10996\\
USA}
\end{aug}

% HISTORY:
\received{\smonth{9} \syear{2009}}
\revised{\smonth{10} \syear{2009}}

% ABSTRACT
\begin{abstract}
How often can we expect a Major League Baseball team
to score at least 20 runs in a single game? Considered a rare event in
baseball, the outcome of scoring at least 20 runs in a game has occurred
224 times during regular season games since 1901 in the American and
National Leagues. Each outcome is modeled as a Poisson process; the time
of occurrence of one of these events does not affect the next future
occurrence. Using various distributions, probabilities of events are
generated, goodness-of-fit tests are conducted, and predictions of
future events are offered. The statistical package R is employed for
analysis.
\end{abstract}

% KEYWORDS

\end{frontmatter}

\section*{Introduction}

In their home opening weekend of the 2009 season, the New York Yankees
and their fans were excited about the future of the Bronx Bombers. The
new ``cathedral'' in Major League Baseball had cost 1.5 billion dollars
to construct. There were certain to be sold-out crowds. The Yankees had
made several off-season signings. Management and fans had expectations
of great pitching and a high-powered offense, capable of scoring LOTS of
runs. That's what Yankees fans wanted in the new ballpark and that's
what they saw; unfortunately, it was the Cleveland Indians who provided
the power. On April 18, 2009, the Tribe scored 14 runs in the second
inning en route to a 22--4 victory over the New York Yankees.
Although teams have scored 20 runs in a game before, this game
established a record for the most runs scored in a 2nd inning (14),
surpassing a record previously held by the Yankees.

This led to a curious question: How often does a Major League team score
20 runs in a game? Can the historical occurrences be used to predict
when the next one might occur? Can the event of scoring twenty or more
runs in a game be classified as a rare event?

\section*{Some scoring history}

In 1901, the American League joined the National League as the second
official league in professional baseball. In that inaugural year for the
new Major Leagues, eight games saw a team score twenty or more runs. The
Boston Americans beat the Philadelphia Athletics 23--12, in a game on
May 2, 1901, in the first occurrence of this feat. The Brooklyn Superbas
accomplished the event twice in 1901. Ninety-eight years later, in 1999,
the feat happened a record nine times.

On August 25, 1922, the Philadelphia Phillies played Chicago at Cubs
Park (later, Wrigley Field). Chicago scored 10 runs in the bottom of the
2nd inning and 14 more runs in the bottom of the 4th. By the end of the
4th inning, the Cubs had a commanding 25--6 lead. The Phillies didn't
give up, fighting back to score 17 runs over the final five frames, but
it wasn't enough. Final score: Chicago 26, Philadelphia 23. This was the
first time each team would score over 20 runs in a game. Chicago didn't
need to bat in the bottom of the 9th, but the two teams combined for 98
at-bats, 21 walks (for a total of 119 plate appearances), 12 doubles, 2
triples, and 3 home runs. And this was the Dead Ball Era. There were 49
runs, 51 total hits, 9 errors committed, and a combination of 25 runners
left on base. Tiny Osborne, a rookie pitcher with the Cubs, picked up
his second save. Cubs leadoff hitter Cliff Heathcote was a perfect
5-for-5, with 2 doubles and 2 walks. He knocked in 4 runs and scored 5
himself. According to the box score, 7000 fans witnessed this slugfest,
which lasted only three hours and one minute.

On June 24, 1950, Pittsburgh was playing in Brooklyn when the two teams
put on a display of batting practice during the game. After 7 innings,
the Dodgers were leading 14--12, but the Bums exploded for 5 runs in
their half of the 8th. The game had to be suspended with 1 out in the
bottom of the 8th inning with the score Dodgers 19, Pirates 12. The game
was completed on August 1, and Brooklyn would add 2 more runs, running
their total to 21.

On May 17, 1979, Philadelphia was again playing the Cubs, this time at
Wrigley Field. After 3-$1/2$ innings, the Phils found themselves with a 17--6 lead (they had scored 7 runs in the top of the 1st, only to allow 6
runs in the bottom of the 1st). Both teams hammered pitching back and
forth, and after 9 innings, the score was tied, at 22--22. The
Phillies scored in the 10th on a Mike Schmidt solo home run. In the
four-hour contest, Schmidt had blasted two home runs, and his Cubs
slugging counterpart, Dave Kingman, hit three. There were 109 at-bats
plus 15 walks, breaking their old record of plate appearances in a game
(with one batter hit by the pitch and two sacrifice flies, 127 men came
to the plate that day). There were 10 doubles, 3 triples, and 11 home
runs. Each team committed two errors. The Phillies avenged the 1922 loss
to the Cubs, as this is only the second game in which two teams each
scored more than 20 runs.

The Cleveland Indians started the 1925 season by beating St. Louis, 21
to 14, to establish the most runs scored by one team (and by both teams)
on Opening Day. 1925 is one of three seasons in which 20 or more runs
were scored in eight different games. The others are 1901 and 1923.

Despite allowing 22 runs to the Indians in 2009, the New York Yankees
have accomplished the rare feat of scoring 20 or more runs in a game the
most times---25. That's not much consolation to having it done to them
early in the season in a new stadium. However, the 1939 Yankees and the
1950 Boston Red Sox are the only two teams to accomplish the feat three
times in the same season. The Sox scored 20 and 29 runs, respectively,
in consecutive games in June 1950. The only other consecutive games log
belongs to the Pittsburgh Pirates, who scored 21 and 24 runs,
respectively, in consecutive games in June 1925. The 29 runs scored by
the Boston Red Sox in 1950 (and equaled by the Chicago White Sox in
1955) stood as a record until 2007, when the Texas Rangers dropped 30
runs on the Baltimore Orioles in the first game of an August 22nd
double-header. Each starting Rangers player had at least two hits. Wes
Littleton was awarded a save in the lopsided victory, since he pitched
three innings in relief of Kason Gabbard, allowing no runs and only two
hits.

Of the 30 teams in Major League Baseball (including their previous teams
since 1901), only Arizona, Houston, and Tampa Bay have not scored 20
runs in a game. For example, the Washington Nationals have not
accomplished the feat, but the Montreal Expos (who became the Nationals
in 2005) did score 21 runs in a 1996 game.

\section*{Modeling the data as a Poisson process with the exponential
distribution}

\begin{figure}[b]

\includegraphics{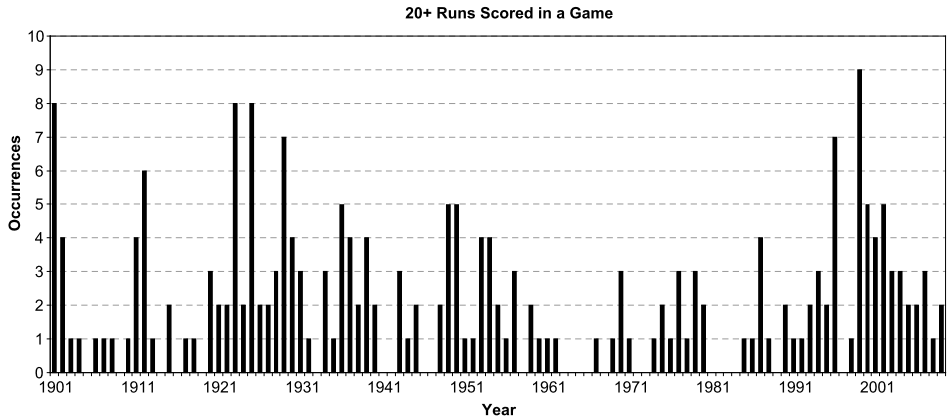}

  \caption{Frequency distribution of 20$+$ run games by season.}\label{fig1}
\end{figure}

In Huber and Glen (\citeyear{HubGle2007}), a rare baseball event is considered to be an
outcome which occurs in fewer than 1\% of all games played. Scoring 20
or more runs in a game falls into this category. Since 1901, when fans
had sixteen teams to root for, a team has scored at least twenty runs
only 224 times. For a frequency distribution by season, see Figure \ref{fig1}.
Given that 171,797 games have been played in the regular seasons and
there were 222 occasions when a team scored 20$+$ runs (from 1901 through
the end of the 2008 season, with two events already in 2009), this
amounts to a percentage of 0.13\%.

%f1 ###

With a value of 0.13\%, this is indeed a rare event that can be modeled
as a Poisson process. The Poisson distribution is a discrete
distribution that is often employed to model the occurrence of events of
a particular type over time. In addition, if the annual number of games
in which at least 20 runs are scored follows the Poisson process, the
exponential distribution can be used to model the times between
consecutive occurrences. We will call these times the ``inter-arrival
times (IATs).'' A random variable $T$ is said to have an exponential
distribution if its probability density function is
\[
f(t;\lambda) = \cases{
\lambda e^{ - \lambda t}, &\quad $t \ge 0$,\cr
0, &\quad otherwise,
}
\]
where $\lambda > 0$. Additionally, if $T$ has an exponential
distribution with parameter $\lambda$, then the expected value of $T$
equals $1/ \lambda$ and the variance of $T$ is equal to $1/\lambda^{2}$.
Both the mean and standard deviation are the same. An important
distinction of the exponential distribution is its ``memoryless''
property. It is well known that the only continuous distribution that
models a memoryless process for inter-arrival times is the exponential
distribution. The time of the last occurrence of one event doesn't
affect the probability of the next occurrence, so for our data, it is
reasonable to assume that these three processes or rare events are
memoryless. For this data, $\lambda$ has an estimated value of 0.00131
or an average IAT of 760.8 games.

In order to develop an exponential model for the data, we needed to
calculate the inter-arrival times between games in which the event
occurred. Using a technique developed by Huber and Glen (2007), we
followed a simple algorithm. Starting with the ``Play Index'' research
tool at \href{http://www.baseball-reference.com}{www.baseball-reference.}
\href{http://www.baseball-reference.com}{com} (accessed May 2009), we generated a list of all
occurrences, sorted by year. However, individual game dates were not
provided. So, we went to \href{http://www.retrosheet.org}{www.retrosheet.org} (accessed May 2009) and investigated
the annual game logs by team. These provided us with verification that
the game actually took place with a team scoring at least 20 runs, and
it gave us the date.

We next counted the total number of games played in the Major Leagues at
the end of the day before an event. For example, the New York Giants
scored 22 runs against the Cincinnati Reds on June 5, 1912. The
standings at the end of the day on June 4th (the previous day) listed
650 total games played. Since it takes two teams to play a game, we
divided 650 by 2 to obtain 325. We then added 1 to that number to get
326. Often, several games started at once on a given day, which makes it
somewhat impossible to ascertain the exact order of games played,
especially in the early part of the 20th century. We therefore assumed
that each game in which the rare event occurred was the first game
played that day (thereby justifying the addition of 1 to the previous
day's total). If the event occurred in the second game of a
double-header, we assumed it was the last game played that day. We
remained consistent with this approach for each game. Continuing with
our example, the New York Giants put up 21 runs against the Boston
Braves on June 20, 1912, two weeks after their first offensive barrage.
This was calculated to be the 425th game of the season. By subtracting
the earlier number from this one, we find that the inter-arrival time
(time between rare events) was $425 - 326 = 99$ games. We used a
continuous count, meaning that we wrapped data from one season to the
next. The Philadelphia Athletics scored 21 runs against the Detroit
Tigers on September 23, 1913, the first event since the last Giants'
game. There were 1232 games played in 1912. The A's game was the 1156th
of the season, giving a difference of 731 games.

\section*{Results}

\subsection*{A single exponential model for the data}

Our preliminary modeling approach was to use all the data and fit an
exponential model (assuming the memoryless property held for 20 run plus
game rare events). The maximum likelihood (ML) estimate for the
exponential model parameter is simply one divided by the sample average.
With an average IAT of 760.8, this results in an estimate of 0.001314444
for the exponential model parameter.

%f2 ###
\begin{figure}[b]

\includegraphics{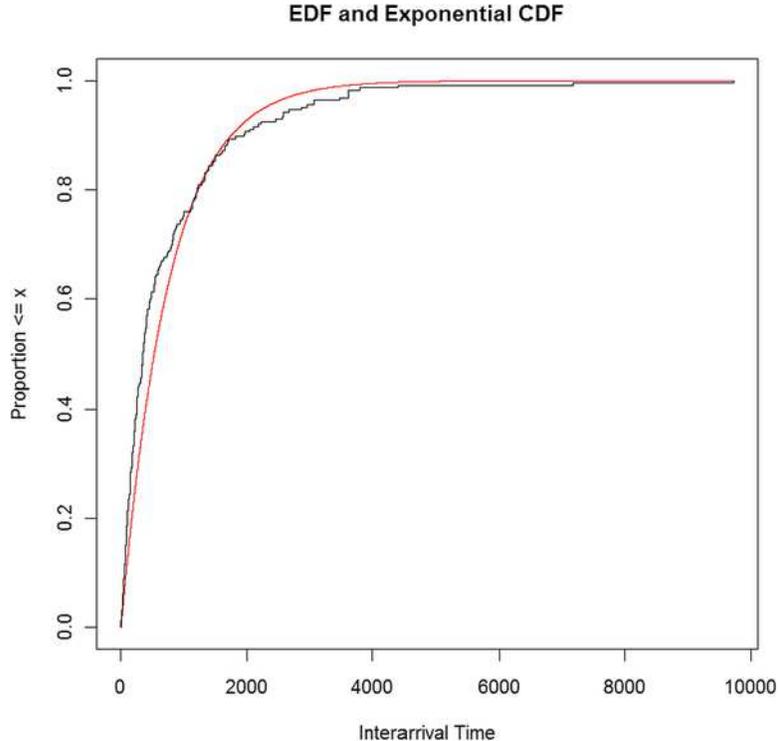}

  \caption{Observed IAT data (EDF) and the fitted exponential model
(CDF).}\label{fig2}
\end{figure}

We next explored how well this model fit the observed data. Initially,
since this is believed to be a memoryless process, we felt that using
the exponential model would be reasonable. However, we observed issues
with the fit throughout much of the distribution. Figure \ref{fig2} shows a
comparison of the exponential model cumulative distribution function
(CDF) using the estimated lambda versus the empirical distribution
function (EDF) for the observed inter-arrival times. At first glance,
the model roughly seems to follow the data. However, there are regions
of the graph where the model does not seem to fit well, which we will
investigate further.

What does the chart tell us? We would expect an 80\% chance of seeing a
team score twenty or more runs in a game within 1200 games played (half
a season). Both curves intersect in this region. Both the EDF and CDF
curves predict a probability of over 90\% for a rare event occurrence
within a season. What about an upcoming weekend, where all teams will
play three games each? That means 45 games total, and we would expect
less than a 10\% chance (closer to 6\%) of seeing an explosion of 20
runs by a single team in a game.

%f3 ###
\begin{figure}

\includegraphics{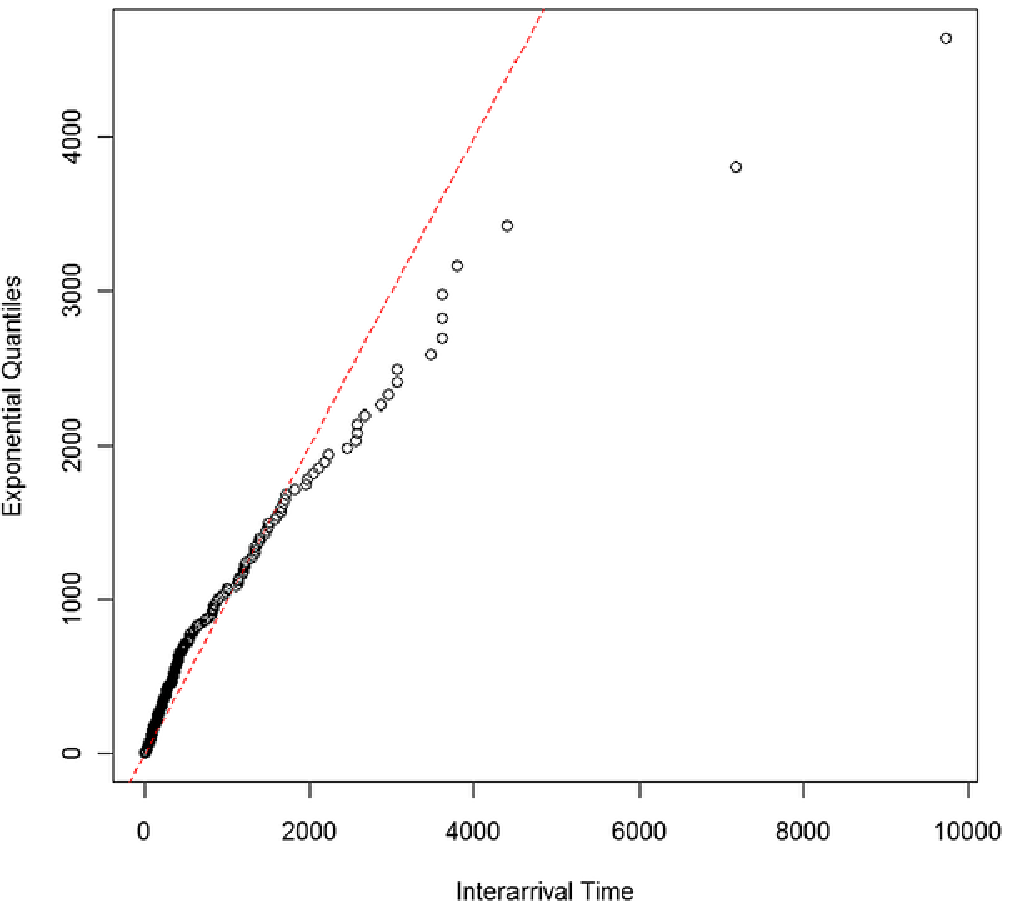}

  \caption{Quantile--quantile (QQ) plot.}\label{fig3}
\end{figure}

To further explore the fit of the model graphically, the best approach
is to use a quantile--quantile (QQ) plot. The QQ plot in Figure \ref{fig3} shows
the observed quantile values on the $x$-axis and the theoretical quantiles
for those values using the exponential distribution on the $y$-axis. If
the distribution fits the data well, the result should be a 45 degree
line---clearly, this distribution is not a good fit. For smaller
inter-arrival times, there are more observed values than one would
expect based upon the exponential model, while with larger times, there
are not as many observed values.

We used three standard goodness-of-fit (GOF) tests to confirm the
conclusions from the graphical analysis of the exponential model. The
three tests used were the Pearson chi-squared ($\chi^{2})$ test (binning
the data by decades, producing a test with 10 degrees of freedom), the
Kolmogorov--Smirnov (K--S) test, and the Anderson--Darling (A--D) test. In
all cases, the null hypothesis is that the model fits the data well.
Results for these three tests are shown in Table \ref{table1}.

%t1 ###
\begin{table}
\tablewidth=180pt
\caption{Goodness-of-fit tests for single exponential model}
\label{table1}
\begin{tabular*}{180pt}{@{\extracolsep{\fill}}ld{2.4}l@{}}
\hline
% Row 1
\textbf{GOF test} & \multicolumn{1}{c}{\textbf{Test statistic}} & \textbf{$\bolds{p}$-value}\\
\hline
% Row 2
K--S test & 0.1581 & $\ll $0.0001\\
% Row 3
A--D test & 8.207 & $\ll $0.0025\\
% Row 4
$\chi^{2}$ test & 44.616  & $\ll $0.05\hphantom{00}\\
\hline
\end{tabular*}
\end{table}

The results for all three tests strongly reject the null hypothesis that
the exponential model fits the data. For the $\chi^{2}$ test, the
critical value for the test at a 5\% significance level is 18.31 and the
test statistic value is well above this threshold. For the A--D test, at
an even more stringent significance level of 0.25\% the critical value
is 2.534; the test statistic again exceeds this value by a large amount.
The K--S statistic exact $p$-value is 0.000026---meaning that there is
little probability that the observed data follows the exponential model
we fit. The GOF tests tend to reject more often for large data (they
have great power to detect departures from the model in such cases).
However, the tests reject very strongly in this case and the graphical
analysis suggests problems with the model as well. We must conclude that
the single exponential model is not a good choice for 20 plus run games.

\subsection*{An improved exponential model based on Era}

To determine the best approach for improving the model, we first
considered the data points with the largest residual (``worst fit'')
values. Due to the nature of the data, these cases are almost certain to
be those with the longest IAT values---the top ten such points shown in
Table~\ref{table2} reflect this fact. However, these games are interesting
as they fall largely (8 of the 10) into two time frames: 1967--1975 and
1985--1992. Looking at the data more closely, we observe that there were
no 20 run games from 1963--1966 and 1981--1984, and generally fewer such
games in the years these 8 data points fall into.

%t2 ###
\begin{table}
\caption{10 worst fit data points using single exponential model}\label{table2}
\begin{tabular}{@{}llc@{}}
\hline
% Row 1
\textbf{Year} & \multicolumn{1}{c}{\textbf{Team}} & \textbf{Games since previous 20$\bolds{+}$ run event}\\
\hline
% Row 2
1985  & Philadephia Phillies & 9723\\
% Row 3
1967 & Chicago Cubs & 7181\\
% Row 4
1974 & Kansas City Royals & 4408\\
% Row 5
1975 & Boston Red Sox & 3608\\
% Row 6
1992 & Milwaukee Brewers & 3471\\
% Row 7
1969 & Oakland A's & 3612\\
% Row 8
1910 & Boston Doves & 3067\\
% Row 9
1948 & St. Louis Cardinals & 3620\\
% Row 10
1986 & Boston Red Sox & 2960\\
% Row 11
1990 & San Francisco Giants & 3795\\
\hline
\end{tabular}
\end{table}

The outlier analysis above suggested that the exponential model may not
be appropriate because the rate of 20 plus games is not constant for the
entire time period of the data. There appear to be periods when the rate
is slower and there are likely other periods where the rate increases.
Scoring and offensive production has changed periodically during
baseball history, leading to the identification of Baseball Eras [taken
from \href{http://www.netshrine.com}{www.netshrine.com} (accessed May 2009)]. These eras are typically defined in
the six periods in Table \ref{table3}.

%t3 ###
\begin{table}[b]
\tablewidth=5.5cm
\caption{Baseball Eras}
\label{table3}
\begin{tabular*}{5.5cm}{@{\extracolsep{\fill}}lc@{}}
\hline
% Row 1
\textbf{Era} & \textbf{Dates}\\
\hline
% Row 2
Dead Ball & 1901--1919\hspace*{7.5pt}\\
% Row 3
Lively Ball & 1920--1941\hspace*{7.5pt}\\
% Row 4
Integration & 1942--1960\hspace*{7.5pt}\\
% Row 5
Expansion & 1961--1976\hspace*{7.5pt} \\
% Row 6
Free Agency & 1977--1993\hspace*{7.5pt}\\
% Row 7
Long Ball  & 1994--current\\
\hline
\end{tabular*}
\end{table}

The years identified in the outlier analysis fall very nicely into the
eras of Expansion and Free Agency. It thus seemed very reasonable to
believe that an appropriate model would use the exponential distribution
but with a different estimated rate parameter for each Baseball Era.
Looking at the data by Baseball Era (Table~\ref{table4}) certainly
supports this approach. We see a noticeable drop in the rate of 20$+$
games in the Free Agency and Expansion Eras, as expected. It does come
as a small surprise that 20$+$ games in the ``Dead Ball'' Era do not occur
at a slower rate; in fact, the average IAT for the era is one of the
shortest. There is also reasonable evidence that exponential models
might be appropriate within eras since the means and standard deviations
are often close (a feature of the exponential distribution). The ``Dead
Ball'' Era has the largest difference in mean and standard deviation. We
will further discuss this era later.

%t4 ###
\begin{table}
\caption{IAT data by Baseball Era}
\label{table4}
\begin{tabular}{@{}lcd{4.4}d{4.4}@{}}
\hline
% Row 1
\textbf{Era} & \textbf{20}$\bolds{+}$ \textbf{games in Era} & \multicolumn{1}{c}{\textbf{Mean IAT}} & \multicolumn{1}{c@{}}{\textbf{SD IAT}}\\
\hline
% Row 2
Dead Ball & 33 & 612.5152 & 859.7318\\
% Row 3
Lively Ball & 68 & 400.7941  & 439.4383 \\
% Row 4
Integration & 37 & 650.1351  & 813.2032 \\
% Row 5
Expansion & 12 & 2307.3333 & 2088.3358\\
% Row 6
Free Agency & 22 & 1561.4091  & 2135.8498 \\
% Row 7
Long Ball  & 53 & 709.7547 & 709.7914\\
\hline
\end{tabular}
\end{table}

\begin{figure}[b]

\includegraphics{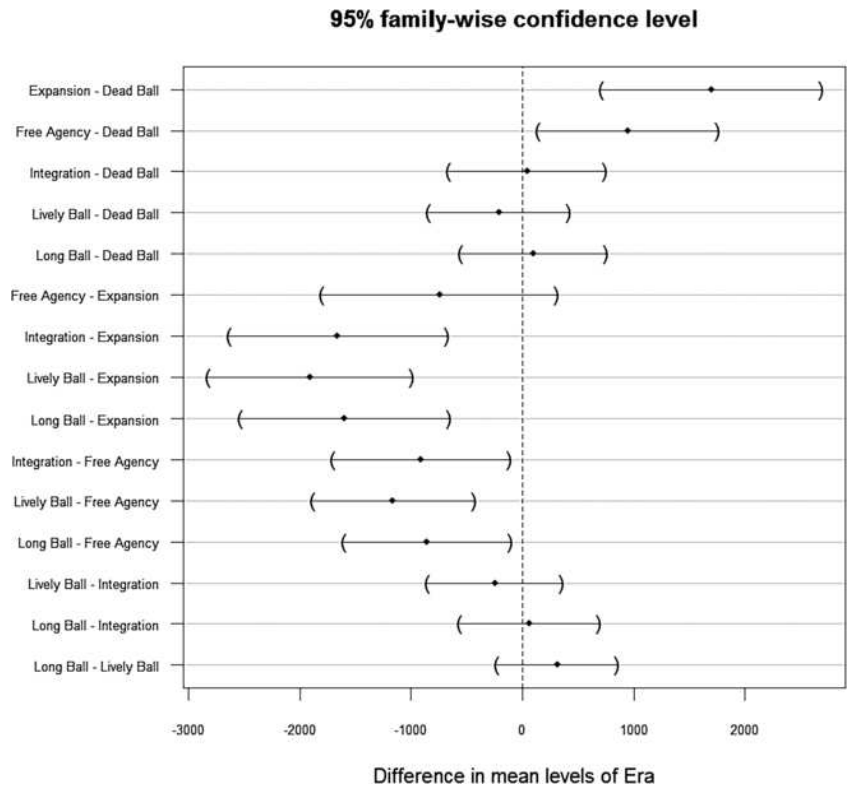}

  \caption{Family-wise confidence intervals for pair-wise comparison of
Baseball Eras.}\label{fig4}
\end{figure}

To confirm the observed differences in eras statistically, we performed
an Analysis of Variance (ANOVA) on the IAT by era. The ANOVA was
statistically significant ($p < 0.0001$) so at least two eras differ
significantly statistically. The pair-wise comparisons (depicted
graphically using the family-wise 95\% confidence intervals in Figure \ref{fig4})
confirm what we observed---the two eras in question (Expansion
and Free Agency) differ from all other eras significantly. They do not,
however, statistically differ from each other, nor do any of the other
eras.

%f4 ###

\begin{figure}[b]

\includegraphics{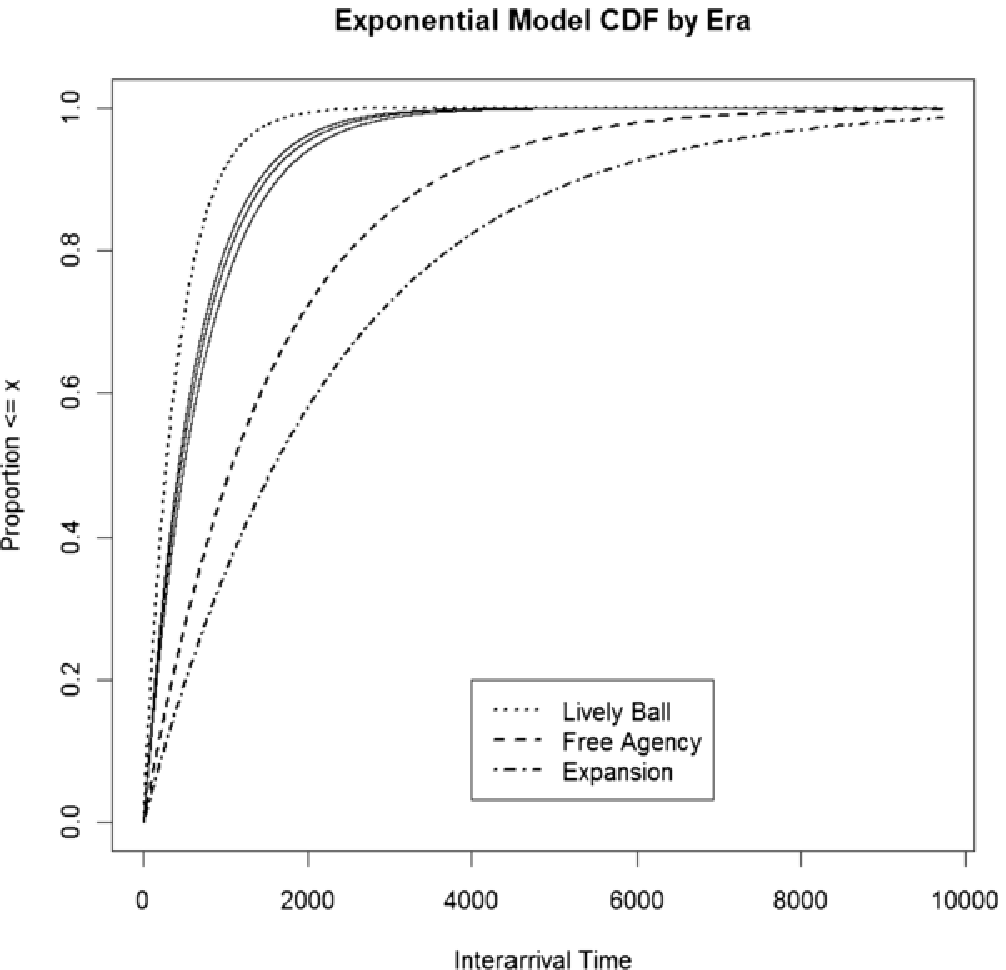}

  \caption{Exponential models by Era.}\label{fig5}
\end{figure}

Fitting exponential models to each Baseball Era, in similar fashion to
the previous modeling efforts, resulted in the estimated rate parameters
in Table \ref{table5}. A~graphical depiction of the resulting models are
shown for each of the six Baseball Eras in Figure \ref{fig5}.

%t5 ###
\begin{table}
\tablewidth=6cm
\caption{ML estimates for exponential rate parameters in each Baseball Era}\label{table5}
\begin{tabular*}{6cm}{@{\extracolsep{\fill}}lc@{}}
\hline
% Row 1
\textbf{Era} & \textbf{Rate parameter}\\
\hline
% Row 2
Dead Ball & 0.001632613\\
% Row 3
Lively Ball & 0.002495138\\
% Row 4
Integration & 0.001538142\\
% Row 5
Expansion & 0.000433401\\
% Row 6
Free Agency & 0.000640466\\
% Row 7
Long Ball  & 0.001408975\\
\hline
\end{tabular*}
\end{table}

%f5 ###

The graphs of the models fit by era dramatically point out the
previously identified eras (Expansion and Free Agency) which have lower
rates of 20 run games. The other eras are more similar, with the
exception of the ``Lively Ball'' Era in which high scoring games occur
at a slightly higher rate.

\begin{figure}

\includegraphics{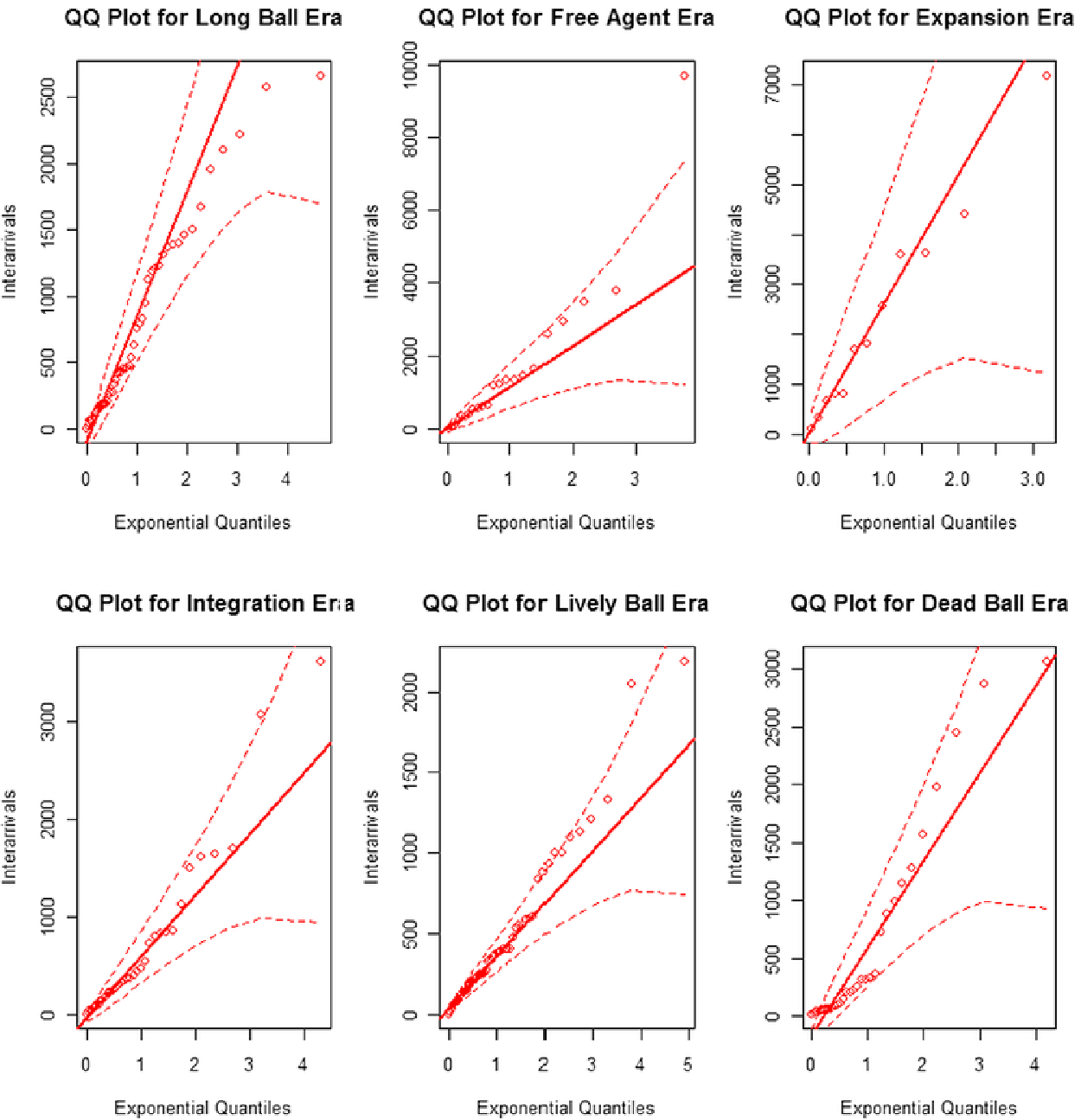}

  \caption{QQ plots for exponential models fit to each Baseball Era.}\label{fig6}
\end{figure}

\begin{figure}

\includegraphics{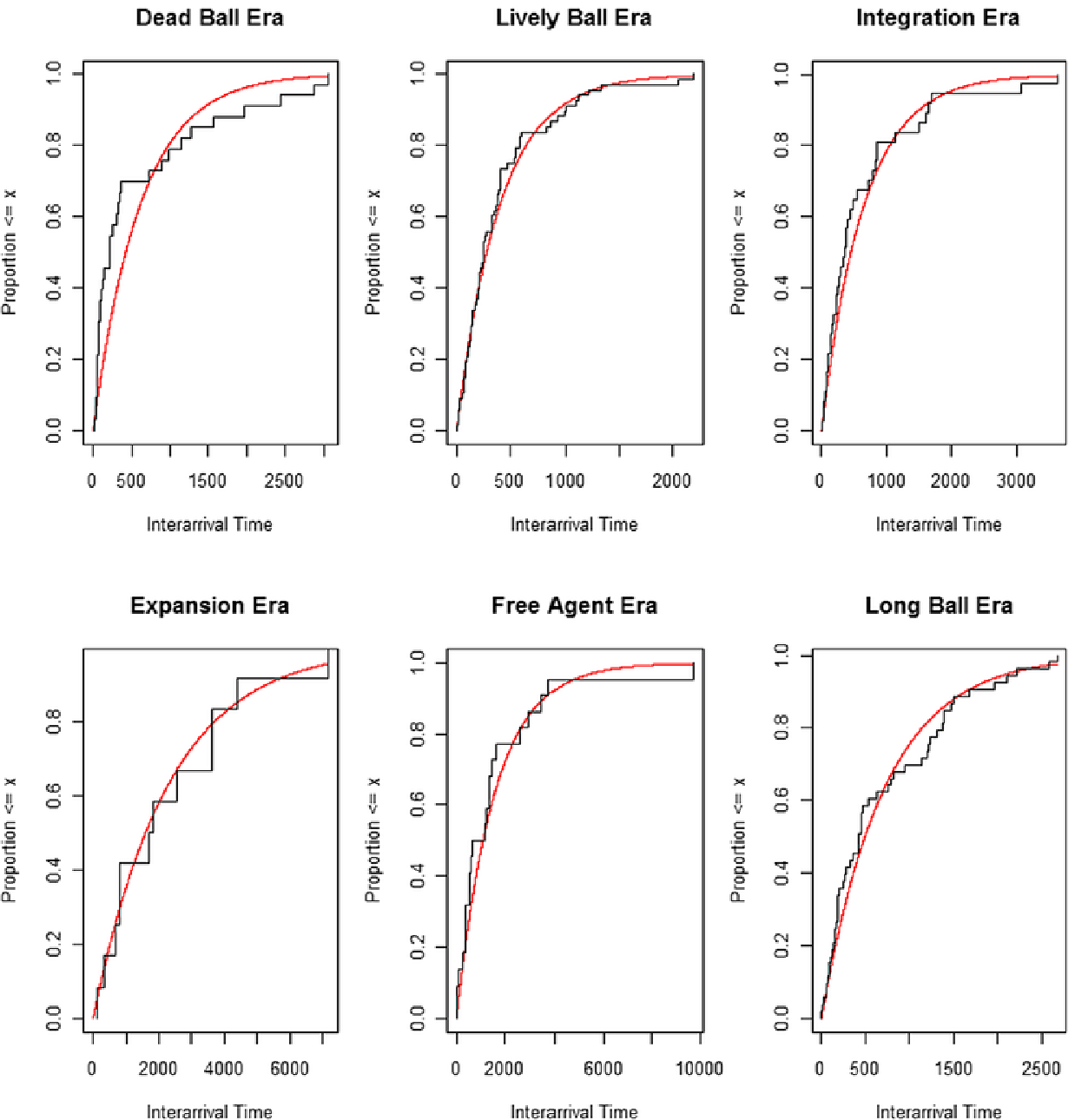}

  \caption{EDF and fitted exponential model CDF for each Baseball Era.}\label{fig7}
\end{figure}

We assess the separate model fits in a similar fashion as we did for the
single exponential model. Graphically, the QQ plots (Figure \ref{fig6}) and the
EDF/CDF plots (Figure \ref{fig7}) suggest that modeling the eras separately has
improved the fit significantly. There are some possible departures from
the model suggested by the QQ plots for a few eras. However, note that
most of the data points are within the 95\% confidence bounds shown on
the graphs. In general, these bounds are much wider for the larger IAT
due to fewer data points---this is also the area where we would expect
more variability and possibly outliers. The more interesting portions of
the plot are those eras where there appear to be more than one ``line''
of points. The ``Dead Ball'' Era is an example. At lower IAT values in
this era the plotted QQ points have a lower slope than the points for
higher IAT's. This suggests the possibility of a missing ``factor'' in
the model.

%f6 ###

%f7 ###

The fitted model CDF's all appear reasonable fits to the observed EDF's
except for the ``Dead Ball'' Era where the model first underestimates the
rate then overestimates.

GOF test results for the six eras are shown in Table \ref{table6}. The K--S
test rejects the null hypothesis of model fit only in the ``Dead Ball''
Era. Similarly, using exponential tables for the A--D test statistics
[D'Agostino and Stephens (\citeyear{AgoSte1986})], only the ``Dead Ball'' Era model fit is questionable. The GOF test
results confirm what we observed graphically---we appear to have well
fit models in all but the ``Dead Ball'' Era.

%t6 ###
\begin{table}
\caption{GOF tests for Baseball Era exponential model fits}
\label{table6}
\begin{tabular}{@{}ld{1.4}cc@{}}
\hline
% Row 1
&& \textbf{A--D approx.}\\
\textbf{Era} & \multicolumn{1}{c}{\textbf{A--D statistic}} &  \textbf{$\bolds{p}$-value} & \textbf{K--S $\bolds{p}$-value}\\
% Row 2
\hline
% Row 3
Dead Ball & 4.054 & $\ll$0.0025 & 0.03348\\
% Row 4
Lively Ball & 0.6172 & $> $0.25\hphantom{00} & 0.5375\hphantom{0}\\
% Row 5
Integration & 0.8282 & 0.15--0.20 & 0.5332\hphantom{0}\\
% Row 6
Expansion & 0.8385 & 0.15--0.20 & 0.9817\hphantom{0}\\
% Row 7
Free Agency & 0.7036 & $>$0.25\hphantom{00} & 0.5431\hphantom{0}\\
% Row 8
Long Ball  & 0.728 & $>$0.25\hphantom{00} & 0.5585\hphantom{0}\\
\hline
\end{tabular}
\end{table}

The modeling issues in the ``Dead Ball'' Era are the result of
inconsistency in certain years during the period. In most years, the
rate of 20$+$ run games is very low in this era, but in 1901--1902 and
then again in 1911--1912 there are an astonishing number of such games.
In 1911 there are 4 twenty run games and in 1912 another 6 for a total
of ten such games in two years. The 1901--1902 seasons have even more
high scoring games recorded with 12 (8 in 1901 alone). This is a very
high rate of these rare events for any era, and particularly surprising
in the ``Dead Ball'' Era where there are a total of 11 other 20$+$ run
games in the remaining 16 years of the era. The increase in 1901--1902
could have been the result of expansion (which often has the effect of
``watering'' down pitching), as the new American League doubled the
number of teams from 8 to 16. The higher rates in these four years
creates the apparent ``dichotomy'' we observed in the graphs of the
model fit and explains the failure of the exponential model for this
era.

\section*{Conclusion}

We model the rare event of games with a team scoring 20 or more runs as
a Poisson process using the exponential distribution to model the IAT
between such games. Our work suggests that the rate of 20$+$ run games is
not constant over time but, rather, the rate varies in Baseball Eras.
Separate models fit to each Baseball Era are successful for eras after
1920. In the ``Dead Ball'' Era prior to 1920 there are issues with the
fitted model due to several years of anomalously high rates of the high
scoring games.

The models fit after 1920 are revealing. During the ``Lively Ball'' Era
beginning in 1920 the rate of 20$+$ run games is the highest of any period
in baseball history, with an average of only 400 games between these
events. The occurrences decrease starting with the ``Integration'' Era
in 1942 when the average number of games between offensive explosions
climbs to 650 games. This agrees with what is generally true in this
era---offensive numbers declined slightly from the ``Lively Ball'' Era.

A further decrease in high scoring games occurs in the two eras to
follow---``Expansion'' and ``Free Agency.'' In fact, from 1961--1976
fans attending baseball games regularly could expect to wait over 2300
games between 20 run outbursts. This, again, agrees with known baseball
history. The ``Expansion'' Era is marked by a decrease in offense as the
league added teams (diluting the talent pool) and, more importantly, the
strike zone was enlarged, giving pitchers a tremendous advantage. The
drop in offense led to some changes in the latter part of the
``Expansion'' Era---fans demanded more scoring. In 1969 the pitching
mound was lowered and in 1973 the American League adopted the designated
hitter. This does raise the rate of 20$+$ games in the ``Free Agency''
Era, although not back to the rates of previous eras. It is not until
the ``Long Ball'' period starting in 1994 that the occurrences of 20$+$
run games increase to levels of the eras before 1960. Somewhat
surprisingly, the rate is still not quite as high as it was then,
despite the tremendous increase in home runs observed in the last
fifteen years.

How can we use this data? Let's explore the probability of predicting a
next occurrence. The Philadelphia Phillies scored 20 runs on June 13,
2008, the second time the World Champions scored 20 runs in that season.
This June game proved to be the last time twenty runs would be scored by
one team in the 2008 campaign. This means that the first game on Opening
Day, 2009, would have an IAT of 1425. According to Figure \ref{fig7} and the Long
Ball Era, an IAT of 1425 would lead to an 87\% chance of an occurrence
of a team scoring 20$+$ runs in a game. Therefore, the Cleveland--New York
game on April 18, 2009, had a high probability of being a game in which
a team scored 20 or more runs. Unfortunately for Yankees fans, it was
the Indians who accomplished the feat. If a month goes by (approximately
300 games), there would be a 34.3\% chance that another event would
occur. This probability increases to 57.4\% after 600 games, given the
rate parameter used.

\printaddresses

\end{document}